# Cross-Cultural Differences in Students' Intention to Use RSS Feeds between Lebanon and the United Kingdom: A Multi-Group Invariance Analysis Based on the Technology Acceptance Model

Ali Tarhini, Department of Computer Science, Brunel University London, UK
Michael James Scott, Department of Computer Science, Brunel University London, UK

Really Simple Syndication (RSS) offers a means for university students to receive timely updates from virtual learning environments. However, despite its utility, only 21% of students surveyed at a Lebanese university claim to have ever used the technology. To investigate whether a cultural influence is affecting intention to use RSS, the survey was extended to the British context to conduct a cross-cultural comparison. Using the Technology Adoption Model (TAM) as a research framework, 437 students responded to a questionnaire containing four constructs: intention to use (INT); attitude towards benefit (ATT); perceived usefulness (PU); and perceived ease of use (PEOU). Principle components analysis (PCA) and structural equation modelling (SEM) were used to explore the psychometric qualities of the scale. The results show that adoption was significantly higher, but also modest, in the British context at 36%. Configural and metric invariance were fully supported, while scalar and factorial invariance were partially supported. Analysis reveals that, as a potential consequence of culture, there are significant differences between PU and PEOU across the two contexts studied, potentially as a consequence of culture. It is recommended that faculty demonstrate to students how RSS can be used effectively in order to increase awareness and emphasise usefulness.

## 1 Introduction

Throughout the last two decades, there has been a profound increase in the use of virtual learning environments, such as Blackboard and Moodle, in higher education institutions to support traditional classroom teaching (Fletcher, 2005; Ngai *et al.*, 2007) and help students meet their educational goals (Clark and Mayer, 2011; O'Neill *et al.*, 2004). However, a lack of portability and pervasiveness in such systems can negatively influence peer interaction, resource acquirement, and content delivery (Cold, 2006). In response to these weaknesses, web-based learning systems have started to integrate Really Simple Syndication (RSS) to provides learners with a means to promptly receive updates using any Internet-enabled device (West *et al.*, 2006). Consequently, enabling learners to be informed about new educational resources in real time, which might include: new teaching materials; reading lists; topics for discussion; or any other course-related announcements. This has been shown to enhance communication among peers (D'Souza, 2006) and help individuals track topics of conversation (Richardson, 2005). RSS has also been used to improve student research by providing access to updated compilations of relevant research references (Asmus *et al.*, 2005). Thus, feeds present one means of providing portability and pervasiveness to virtual learning environments in a way that facilitates collaboration and the dissemination of new information.

However, while RSS feeds are used successfully in many organizations, the use of RSS in education can entail the problem of students' low level of usage (Cold, 2006). Despite its utility and widespread deployment, a Lebanese institution has encountered a high level of resistance to the system. So, despite having the potential to enhance learning through student interaction, a low level of acceptance has meant its benefits have not been fully realised (Saadé and Bahli, 2005; Kim and Moore, 2005). It is therefore imperative for practitioners and policy makers to understand the acceptance of learning systems in the Lebanese context in order to increase usage, and thereby enhance the learning opportunities available to students in Lebanon. In particular, focusing on whether cultural or socio-economic influences are affecting students' attitudes towards e-learning tools.

Various models and theories have been developed to investigate and understand and predict the acceptance of technology. Examples include: the Technology Acceptance Model (TAM) (Davis, 1989; Davis *et al.*, 1989); the Theory of Reasoned Action (TRA) (Ajzen and Fishbein, 1980; Fishbein and Ajzen, 1975); Innovation Diffusion Theory (IDT) (Rogers, 1995); the Theory of Planned Behaviour (TPB) (Ajzen, 1991); and the Unified Theory of Acceptance and Use Technology (UTAUT) (Venkatesh *et al.*, 2003). This research employs TAM in order to understand and explain the relationship between individuals' perceptions and behavioural intentions towards RSS. This is because TAM has been widely used in similar information systems research; see (Yousafzai *et al.*, 2007); due to its parsimonious structure and acceptable explanatory power (Venkatesh and Bala, 2008). Furthermore, the validity and reliability of TAM across a number of different technologies and usage contexts have been examined (Teo and Noyes, 2011; Park, 2009; Venkatesh and Davis, 2000).

A criticism of TAM is that it can be affected by biases in cross-cultural contexts (McCoy *et al.*, 2005; Teo *et al.*, 2008; Straub *et al.*, 1997). For example, Straub et al. (1997) tested TAM across three different cultures, finding that TAM produces different explanatory power in behavioural intention between Japan and the United States (i.e only 1%), but similar power between the United States and Switzerland (i.e 10%). However, the argument that TAM doesn't serve equally across cultures is inconsistent in the prior literature (McCoy *et al.*, 2005; Zakour, 2004; Srite and Karahanna, 2006). Of particular concern, in this case, is whether TAM will be suitable for use in the Lebanese context, and whether it can be used to compare Lebanese culture to that of other nations. This is because TAM has not been widely tested in developing countries (Teo *et al.*, 2008). Consequently, there is a gap in the literature, and so it is important to first test the appropriateness of TAM through exploring the psychometric properties of the research instrument to ensure measurement invariance, as well as adequate reliability and validity.

The authors hypothesise that there may be cultural influences that are affecting technology adoption in Lebanon. Thus, this study also proposes to compare a Lebanese sample to a sample from a different culture to explore the differences. The United Kingdom was selected as an example of a typical developed country that could be used to conduct such a cross-cultural comparison. This country was chosen because, as shown below in Table 1, the United Kingdom and Lebanon represent nearly reverse positions on all of Hofstede's (2005) cultural dimensions. In addition, the investment in technology in the educational system is still immature in comparison since universities and higher education institutions in Lebanon support traditional styles of pedagogy due to the lack of financial support and trained staff (Baroud and Abouchedid, 2010).

**Table 1**
Differences between Britain and Lebanon on Hofstede's (2005) Cultural Dimensions

| Country | Power Distance | Masculinity | Individualism | Uncertainty Avoidance |
|---|---|---|---|---|
| Lebanon | 80 | 53 | 38 | 68 |
| United Kingdom | 35 | 66 | 89 | 35 |

Consequently, this study extends the literature by applying TAM to Lebanon to examine differences in factors that may affect the acceptance and adoption of RSS feeds among British and Lebanese students. The results of this study would be of interest to the research community since it explores the generalizability and validity of TAM in a cross-cultural context in order to explore differences in its applicability in the context of e-learning. This will help policy makers and practitioners to gain a deeper understanding of the students' acceptance of e-learning technology and consequently lead to enhancements in technology acceptance and learning.

## 1.1 The Technology Acceptance Model (TAM)

Davis (1989) developed the technology acceptance model (TAM) through the theoretical foundation for the Theory of Reasoned Action (TRA) (Malhotra and Galletta, 1999). TRA is a model pertaining to social psychology concerned with the specifics of intended behaviours (Ajzen and Fishbein, 1980). TRA posits that an individual's behaviour and intent to behave is a function of that individual's attitude toward the behaviour and their perspectives regarding the behaviour. Behavioural intention also is determined via subjective norms, as behaviour results as a function of all attitudes and beliefs (Masrom, 2007).

TAM aims "to provide an explanation of the determinants of computer acceptance that is general, capable of explaining user behaviour across a broad range of end-user computing technologies and user populations, while at the same time being both parsimonious and theoretically justified" (Davis, 1989). According to Venkatesh et al (2003), TAM presumes that behavioural intention is usually formed as a result of conscious decision making processes. The three variables (see Figure 1), perceived usefulness (PU), perceived ease of use (PEOU), and attitude (ATT) are keys to predicting a user's perceptions towards usage and acceptance of technology (Davis, 1989). The TAM postulates that PEOU and PU predict the user's attitude towards the system (ATU), behavioural intention (BI) is predicted by the user's attitude (A), and the actual use of the system is predicted by BI. Furthermore, the PEOU has a significant influence on PU.

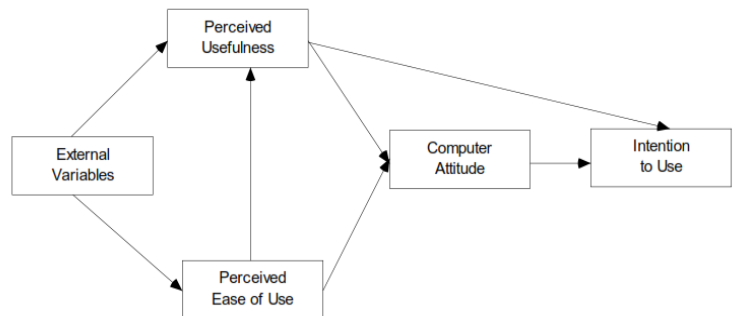

**Fig 1.** Research Model based on Davis, Bagozzi, & Warshaw's (1989) Technology Acceptance Model (TAM)

The main foundation of our research is based on TAM and drawing from previous literature that used TAM in an educational context in order to reflect the usage and acceptance of RSS in education. The overall conceptual model is illustrated above in Figure 1 and the sections which follow explain and justify each of the predicted relationships in light of previous findings from the literature.

### 1.2  Perceived ease of use (PEOU) and Perceived Usefulness (PU)

Perceived usefulness (PU) is a predictor that measures individuals' beliefs regarding whether the use of a particular technology system will improve her or his performance (Davis *et al.*, 1989). Perceived usefulness was used in this study to investigate students' beliefs about obtaining benefits regarding the use of Blackboard's system as well as to predict students' beliefs of using RSS on the Blackboard system. The selection of this factor in the research model was due to the direct and significant influence on user's attitude to use the technology and also behavioural intention to use the system, which comes from the previous studies (e.g., Teo et al., 2008). Perceived ease of use (PEOU) is a predictor that measures an individual's beliefs regarding the use of a particular technology system free of effort (Davis, 1989). The PEOU construct was selected in order to investigate students' attitudes regarding using Blackboard's system free of effort, as well as to predict students' intentions of using RSS on the Blackboard system.

TAM posits that PEOU and PU predict the user's attitude towards the system (ATU). As such, it is expected that users with high level of PU are more likely to have positive attitudes about using the technology. Similarly users with high level of PEOU are also expected to induce positive attitudes. Furthermore, according to Davis (1989), PU was found to mediate the effect of PEOU on attitude. In another words, PEOU indirectly has an impact on attitudes through PU.

### 1.3  Attitude toward using (ATU)

Attitude toward using (ATU) is a predictor that investigates individuals' beliefs regarding using a particular technology. The casual relationship between PU, PEOU and ATT towards using the technology is supported considerable number of studies e.g. (Cheng *et al.*, 2006; Yu *et al.*, 2005; Tarhini *et al.*, 2013). Furthermore, Gao (2005) indicated in his study 'Educational Hypermedia' that attitude toward using had a direct and significant effect on intention to use. Also, Malhotra and Galletta (1999) claimed in their study that attitude toward using had a direct and significant effect on intention to use.

However, attitude toward using has varying degrees of effectiveness based on the field of study, sample, or techniques used for analysis. On the other hand, Masrom (2007), found in her study of 'learning online' that attitude toward using did not have a direct and significant effect on intention to use. There are differences in significant and insignificant effects of the intention to use based on the field or sample of the study. This is because the term 'attitude' can be interpreted quite broadly and could be directed towards many different facets of using a system, such as 'attitude towards features', 'attitude towards purpose' or 'attitude towards benefit'; of which, the latter is the focus in this article.

### 1.4 Behavioural Intention (BI) to use the system

The presence of behavioural intention (BI) in the TAM is one of the major differences with TRA. BI is considered to be an immediate antecedent of usage behaviour and gives an indication about an individuals' readiness to perform a specific behaviour. In TAM, both PU and PEOU influence an individual's intention to use the technology, which in turns influence the usage behaviour (Davis, 1989). There is significant supports in the literature for the relationship between PU, PEOU and ATT on BI (Venkatesh and Davis, 2000; Taylor and Todd, 1995; Davis *et al.*, 1989), particularly in the context of e-learning studies (Zhang *et al.*, 2008; Yi-Cheng *et al.*, 2007; Park, 2009; Saeed and Abdinnour-Helm, 2008; Liu *et al.*, 2010). It is worth noting, however, that actual usage (AU) of the system was excluded from this study because it was found to be challenging to track individual users based on the available server system logs. This is because RSS feeds are available without requiring a login, to facilitate ease of access. Thus, it was impossible to distinguish individual users, or even distinguish between individual mobile devices, with the data available. Therefore, it was deemed appropriate to only measure behavioural intention.

### 1.5 Aim of the Study and Hypotheses

The overall research question addresses whether intention to use RSS feeds in the higher education setting is influenced by cultural factors. This will be addressed by examining five hypotheses based on TAM, as shown below in Figure 2.

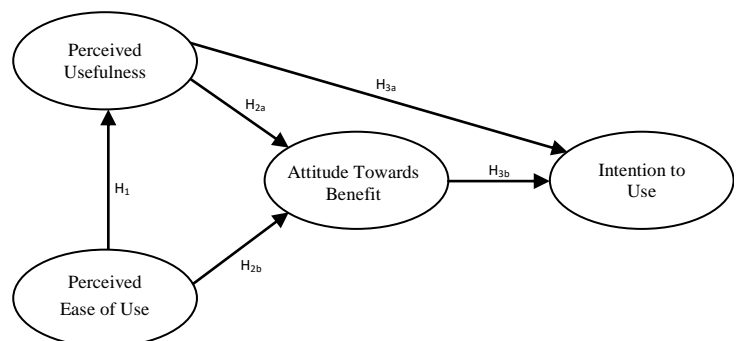

**Fig. 2.** Hypotheses based on the Technology Acceptance Model (Davis et al, 1989)

$H_1$: Students' perceived ease of using RSS feeds (PEOU) will significantly influence the perceived usefulness of RSS feeds (PU) in both the Lebanese and British contexts, equally.

$H_{2a}$: Students' perceived usefulness of RSS feeds (PU) will significantly influence attitude towards the benefits of using RSS (ATT) in both the Lebanese and British contexts, equally.

$H_{2b}$: Students' perceived ease of using RSS feeds (PEOU) will significantly influence attitude towards the benefits of using RSS (ATT) in both the Lebanese and British contexts, equally.

$H_{3a}$: Students' perceived usefulness of RSS feeds (PU) will significantly influence intention to use RSS feeds available on Blackboard Learn (INT) in both the Lebanese and British contexts, equally.

$H_{3b}$: Students' attitude towards the benefits of using RSS (ATT) will significantly influence intention to use RSS feeds available on Blackboard Learn (INT) in both the Lebanese and British contexts, equally.

Further to these hypotheses, the following are also made:

| | |
|---|---|
| $H_4$: | Students' experience with using RSS feeds will be significantly higher in the United Kingdom, compared to Lebanon. |
| $H_{5a}$: | Students' mean perceived usefulness of RSS feeds (PU) will be significantly higher in the United Kingdom, compared to Lebanon. |
| $H_{5b}$: | Students' mean perceived ease of using RSS feeds (PEOU) will be significantly higher in the United Kingdom, compared to Lebanon. |
| $H_{5c}$: | Students' mean attitude towards the benefits of using RSS (ATT) will be significantly higher in the United Kingdom, compared to Lebanon. |
| $H_{5d}$: | Students' mean intention to use RSS feeds available on Blackboard Learn (INT) will be significantly higher in the United Kingdom, compared to Lebanon. |

## 2   Methodology

Consistent with previous empirical research in technology acceptance e.g. (Venkatesh and Bala, 2008; Venkatesh *et al.*, 2003) and similar work within the e-learning context e.g. (Zhang *et al.*, 2008; Liaw, 2008), a quantitative approach was adopted. A 25-item questionnaire was administered to students at an institution in the United Kingdom and an institution in Lebanon, by a process of convenience sampling. The questionnaire contained at least five items for each of the proposed constructs in the TAM model (PU, PEOU, ATT and INT). These items were adapted for the context of using new RSS feeds that had been introduced within the Blackboard Learn virtual learning environment.

Data collected from this survey were analysed using principle components analysis (PCA) in SPSS 20.0.0 and structural equation modelling (SEM) using AMOS 18.0.1. The PCA technique was applied to cull the larger set of items down to a smaller, more parsimonious scale containing items that were likely to be invariant across multiple groups, while also ensuring that the proposed factor structure was appropriate for the items included in the scale. The SEM technique was then applied to ensure that adequate construct validity was present in the data and to verify the level of measurement invariance those items achieved.

As Straub (1997) points out, it is important that any hypothesised latent constructs are measured in an appropriate manner. Researchers must ensure that they are actually measuring what they believe to be measuring by ensuring that an appropriate level of construct validity is found. Hair et al (2010) show that if adequate face validity, convergent validity and discriminant validity are found, then together these present sufficient evidence for construct validity. That is, participants understand the meaning of every item in the scale (face validity), a set of items expected to measure a particular latent factor converge on that factor with strong factor loadings (convergent validity), and the extent to which constructs differ by not sharing variance can be established (discriminant validity).

Just as, from the literature, researchers conduct such multi-group analyses to compare a treatment group with a reference or baseline group, such multi-group comparisons allow researchers to determine whether the differences between two groups, those being the independent variables, are affecting some dependent variables. In this case, the analyses were performed to compare whether cultural factors interact with intention to use RSS feeds, based on studying two culturally distinct groups: university students in Lebanon; and university students in the United Kingdom. However, before the results of such a cross-cultural comparison can be interpreted meaningfully, assumptions of measurement invariance need to first be verified. This is because, based on a review of the literature, Vandenberg and Lance (2000) emphasise that at least some configural, metric, scalar and factorial invariance should be established.

Such tests of measurement and structural invariance generally fall into one of five questions about how participants interpret items in an instrument (Byrne, 2006): (1) do the items that comprise an instrument operate in a similar fashion across groups; (2) are the constructs and factor structure equivalent across groups; (3) is the causal structure of the constructs the same across groups; (4) are the means of the factor scores invariant across the two groups; and (5) does the factorial structure of an instrument replicate across different independent samples of the same population? Once such questions have been answered, researchers can have confidence that the meaning of responses to particular items in a scale do not differ significantly across multiple groups and are reliable within-groups. Thus, as recommended by Vandenberg and Lance (2000), tests were performed to ensure that the configuration of factors were the same across the two cultures (configural invariance), rating scales were interpreted similarly (metric invariance), the quantifiable meanings of the scales meant the same to participants from both cultures

(scalar invariance), and factor variances are homogenous indicating the equality of relationships between the latent factors (factorial invariance).

Once invariance has been established in the measurement model, the structural model can be tested to examine the relationships between the constructs. The differences between the model structure in the Lebanese and British samples were compared using z-tests on the correlation coefficients between pairs of constructs. Estimated factor scores were also generated using data imputation in AMOS using the regression method. The resulting data from Lebanon and the UK were subsequently compared using independent sample t-tests in SPSS.

## 2.1　Participants and Sampling

The participants in this study comprised of 202 students attending a university in Britain and 235 students attending a university in Lebanon. All participants were studying in an English-language setting and were assumed to be computer-literate, as both institutions were predominantly running courses in engineering, technology and ICT. Both institutions were also making extensive use of the Blackboard Learn virtual learning environment.

No course credit or other rewards were given to participants in this study. Prior to completing the questionnaire, all participants were briefed on the purpose of the work, and their right to choose not to participate. On average, each participant took no more than 10 minutes to complete the questionnaire. Details of the participants are shown below in Table 2.

**Table 2**
Demographic Information of Participants

| Country | Age | | Gender | | Education | | Experience with RSS | |
|---|---|---|---|---|---|---|---|---|
| | *M* | *SD* | *Male* | *Female* | *Undergrad* | *Graduate* | *Yes* | *No* |
| Lebanon | 22.6 | 4.4 | 121 | 114 | 102 | 133 | 51 | 184 |
| United Kingdom | 21.8 | 4.9 | 91 | 111 | 101 | 101 | 74 | 128 |
| *Group Differences* | t | 1.688 | $\chi^2$ | 1.804 | $\chi^2$ | 1.900 | $\chi^2$ | 11.859 |
| | *p* | .072 | *p* | .179 | *p* | .168 | *p* | .001 |
| | | | | | | | $H_4$ Result: | Supported |

*Notes:* M = mean; SD = standard deviation.

The sample was collected using a non-probabilistic, self-selection method, and should therefore be considered a convenience sample. More specifically, the empirical data were collected by means of self-administrated questionnaire containing 21 questions. The survey was conducted in-person across a period of 3-weeks by two researchers moving to multiple locations within each institution, namely the libraries, computer suites and study areas.

## 2.2　Measures

The instrument was administered in English to all of the students who volunteered to participate. The questionnaire was first pre-tested for content and face validity in both settings. Apart from providing demographic their demographic information, they responded to 25 items, adapted from the work of Davis (1989), including: perceived usefulness of the RSS feeds (PU) (5 items); perceived ease of using RSS feeds (PEOU) (6 items); attitudes towards the potential benefits of using RSS (ATT) (8 items); and intention to use particular RSS feeds on Blackboard Learn (INT) (6 items). These items were measured using a 5-point Likert scale ranged from strongly disagree to strongly agree.

## 3　Results

The results are presented initially focusing on how the research instrument was refined, based on the results of a principle components analysis and matrix independence tests using Fisher's method, and descriptive statistics for are shown to indicate general responses to indicate the general responses to the items representing the constructs being measured. Following this, the results of a series of measurement invariance tests based on a structural equation modelling technique are shown; determining whether the cultures studied can be meaningfully compared using the refined instrument. Subsequently, a series of t-tests and z-tests test whether the data supports or does not support the hypotheses presented.

### 3.1 Instrument Development and Refinement

The instrument was refined based on a principle component analysis of the original 25 items included in the survey. Items with low loadings (< .4) on their theorised component, significant cross loadings (> .4 in a different component), and items belonging to undefined components were removed. An analysis using Fisher's method on these items showed that the two rotated component loading matrices (for Lebanon and the United Kingdom) were significantly different from each other ($X^2 = 303.79$, $df = 190$, $p < .001$) and so items with significant z-scores were also removed, maintaining at least three items per factor, until an adequate solution was found ($X^2 = 105.81$, $df = 96$, $p = .231$), as shown in Table 3. This refined scale contained 12 of the original 25 items, with a KMO of .763 and a significant Bartlett's indicating adequate factorability. Both Catell's scree plot criterion and Kaiser's eigenvvalue criterion indicated the 4 component solution, as hypothesised, were appropriate. The overall variance explained for the two models were 73% and 71%, for the British and Lebanese samples respectively.

**Table 3**
Principle Component Analysis and Fisher's Test of Independence

|  | British Sample ($n = 202$) | | | | Lebanese Sample ($n = 235$) | | | | z-Tests | | | |
|---|---|---|---|---|---|---|---|---|---|---|---|---|
|  | INT | PEOU | PU | ATT | INT | PEOU | PU | ATT | INT | PEOU | PU | ATT |
| INT1 | .892 | .151 | .064 | .171 | .844 | .070 | .044 | .178 | 2.02 | 0.84 | 0.20 | -0.06 |
| INT2 | .900 | .059 | .145 | .210 | .862 | .000 | .133 | .103 | 1.75 | 0.60 | 0.12 | 1.14 |
| INT3 | .898 | .057 | .116 | .113 | .758 | .004 | .126 | .238 | 4.88** | 0.54 | -0.10 | -1.33 |
| PEOU1 | .101 | .797 | .181 | .183 | .162 | .811 | .025 | .033 | -0.64 | -0.40 | 1.64 | 1.56 |
| PEOU2 | .074 | .888 | -.026 | .047 | .003 | .859 | .000 | .079 | 0.73 | 1.25 | -0.27 | -0.33 |
| PEOU4 | .067 | .783 | .214 | .004 | -.083 | .772 | .073 | .006 | 1.54 | 0.27 | 1.49 | -0.02 |
| PU1 | .116 | .059 | .796 | .201 | .092 | .004 | .858 | .112 | 0.26 | 0.56 | -2.07 | 0.94 |
| PU2 | .093 | .085 | .868 | .112 | .152 | -.024 | .890 | .073 | -0.62 | 1.13 | -1.01 | 0.41 |
| PU3 | .086 | .206 | .708 | .113 | .047 | .112 | .689 | .128 | 0.40 | 1.00 | 0.38 | -0.16 |
| ATT6 | .214 | -.025 | .106 | .738 | .187 | -.001 | .184 | .798 | 0.29 | -0.25 | -0.82 | -1.52 |
| ATT7 | .084 | .207 | .119 | .864 | .302 | .038 | .077 | .839 | -2.35* | 1.78 | 0.43 | 0.92 |
| ATT8 | .153 | .059 | .202 | .728 | .073 | .090 | .092 | .847 | 0.83 | -0.32 | 1.16 | -3.34** |
|  |  |  |  |  |  |  |  |  |  |  |  |  |
| Eigenvalue | 4.133 | 1.855 | 1.545 | 1.224 | 3.636 | 1.965 | 1.652 | 1.291 |  |  | Fishers $X^2$ | 105.81 |
| % VE | .344 | .157 | .128 | .101 | .302 | .163 | .137 | .107 |  |  | $p$ | 0.231 |

*Notes:* Factor loadings have been rotated using a varimax rotation with Kaiser normalization; %VE = percentage of variance extracted.
* $p < .05$, ** $p < .01$

Following this stage of item refinement, the following 12 items were included in subsequent analyses, as described below in Table 4. The descriptive statistics for each item are also shown in Table 5.

**Table 4**
List of Constructs and Corresponding Items in Final Scale for Further Analysis

| Construct | Item | Description |
|---|---|---|
| Perceived Usefulness of RSS Feeds (PU) | PU1 | I would like to be informed about any activities on Blackboard |
|  | PU2 | I would like to receive updates on Blackboard as soon as published |
|  | PU3 | I would like to receive course information daily |
| Perceived Ease of Using RSS Feeds (PEOU) | PEOU1 | I find using RSS feeds on Blackboard Learn easy to use |
| | PEOU2 | I find it easy to check for information regarding my courses with RSS |
| | PEOU4 | I find it easy to look up all recently uploaded materials using RSS |
| Attitude Towards Potential Benefit (ATT) | ATT6 | I could improve my learning performance by receiving new information |
| | ATT7 | I could enhance my learning skills |
| | ATT8 | I could increase my interaction with Blackboard |
| Intention to Use RSS Feeds (INT) | INT1 | I intend to receive information through the "course content" feed |
| | INT2 | I intend to receive information through the "announcement" feed |
| | INT3 | I intend to receive information through the "discussion" feed |

**Table 5**
Mean, Standard Deviation, Skewness and Kurtosis of Scale Items

| | Pooled Sample (*n* = 437) | | | | British Sample (*n* = 202) | | | | Lebanese Sample (*n* = 235) | | | |
|---|---|---|---|---|---|---|---|---|---|---|---|---|
| | M | SD | Sk | K | M | SD | Sk | K | M | SD | Sk | K |
| INT1 | 3.73 | .989 | -.742 | .398 | 3.70 | .973 | -.848 | .421 | 3.75 | 1.004 | -.742 | .398 |
| INT2 | 3.69 | .979 | -.652 | .159 | 3.75 | .947 | -.719 | .406 | 3.64 | 1.005 | -.652 | .159 |
| INT3 | 3.46 | 1.007 | -.183 | -.183 | 3.73 | .946 | -.817 | .480 | 3.23 | 1.002 | -.183 | -.183 |
| PEOU1 | 4.09 | .895 | -.753 | -.250 | 3.75 | .987 | -.490 | -.477 | 4.39 | .685 | -.753 | -.250 |
| PEOU2 | 3.93 | .910 | -.636 | -.303 | 3.72 | .942 | -.428 | -.344 | 4.11 | .843 | -.636 | -.303 |
| PEOU4 | 3.97 | .849 | -.596 | .061 | 3.69 | .906 | -.367 | -.227 | 4.20 | .721 | -.596 | .061 |
| PU1 | 3.89 | .826 | -.361 | .075 | 3.82 | .885 | -.772 | .839 | 3.94 | .769 | -.361 | .075 |
| PU2 | 3.76 | .805 | -.184 | -.070 | 3.76 | .836 | -.559 | .411 | 3.75 | .780 | -.184 | -.070 |
| PU3 | 3.51 | .877 | -.276 | .245 | 3.61 | .864 | -.287 | -.089 | 3.43 | .881 | -.276 | .245 |
| ATT6 | 3.66 | .977 | -.362 | -.527 | 3.75 | .869 | -.141 | -.727 | 3.58 | 1.057 | -.362 | -.527 |
| ATT7 | 3.94 | .980 | -.673 | -.141 | 4.12 | .856 | -.809 | .335 | 3.79 | 1.053 | -.673 | -.141 |
| ATT8 | 3.57 | 1.044 | -.219 | -.722 | 3.73 | .936 | -.314 | -.586 | 3.44 | 1.113 | -.219 | -.722 |

*Notes:* M = mean; SD = standard deviation; Sk = skewness; k = kurtosis; n = sample size, after removing outliers and invalid responses

The descriptive statistics presented above in Table 5 indicate a somewhat positive disposition towards RSS feeds. All means were greater than the midpoint (2.5) for both samples, ranging from 3.51 to 4.09. While the standard deviation (SD) values ranged from .685 to 1.113 for the Lebanese sample, indicating greater variability compared to .836 and .973 in the British sample, these values could still be considered a narrow spread around the mean. However, to ensure adequate multivariate normality in the sample, several cases were removed as outliers based on having a Mahalanobis distance greater than 35 from the centroid.

As the maximum-likelihood estimation method was applied during the evaluation of the structural equation model, it is important that the distribution of the data does not significantly depart from a multivariate normal distribution. This can be verified through examination of the univariate distribution index values, with skew indices greater than 3.0 and kurtosis indices greater than 10 indicative of severe non-normality (Kline, 2005). Since the values of the variables for both samples fall well within the guidelines, therefore the data in this study were considered to be normal.

### 3.2    Examination of reliability, convergent validity and discriminant validity

The next step is to assess convergent validity, discriminant validity in addition to reliability in order to evaluate that the psychometric properties of the measurement model are adequate. As advocated by Hair et al (2010), this can be established in terms of composite reliability (CR), average variance extracted

(AVE). The results are shown below in Tables 6.

**Table 6**
Convergent and Discriminant Validities[a]

|  | Pooled Sample (*n* = 437) | | | | | British Sample (*n* = 202) | | | | | Lebanese Sample (*n* = 235) | | | | |
| --- | --- | --- | --- | --- | --- | --- | --- | --- | --- | --- | --- | --- | --- | --- | --- |
|  | FL | CR | AVE | MSV | ASV | FL | CR | AVE | MSV | ASV | FL | CR | AVE | MSV | ASV |
| PU1 | .756 | | | | | .736 | | | | | .792 | | | | |
| PU2 | .885 | .780 | .551 | .105 | .078 | .832 | .771 | .533 | .179 | .133 | .930 | .796 | .579 | .092 | .056 |
| PU3 | .545 | | | | | .603 | | | | | .503 | | | | |
| PEOU1 | .770 | | | | | .780 | | | | | .697 | | | | |
| PEOU2 | .804 | .801 | .575 | .034 | .023 | .778 | .796 | .566 | .118 | .098 | .844 | .759 | .518 | .018 | .010 |
| PEOU4 | .696 | | | | | .697 | | | | | .596 | | | | |
| ATT6 | .694 | | | | | .617 | | | | | .723 | | | | |
| ATT7 | .891 | .807 | .586 | .224 | .116 | .866 | .758 | .517 | .179 | .154 | .910 | .759 | .616 | .279 | .123 |
| ATT8 | .695 | | | | | .649 | | | | | .706 | | | | |
| INT1 | .812 | | | | | .865 | | | | | .792 | | | | |
| INT2 | .870 | .853 | .660 | .224 | .112 | .932 | .913 | .779 | .167 | .110 | .797 | .809 | .586 | .279 | .123 |
| INT3 | .750 | | | | | .848 | | | | | .705 | | | | |
|  | PU | INT | PEOU | ATT | | PU | INT | PEOU | ATT | | PU | INT | PEOU | ATT | |
| PU | (.742) | | | | | (.730) | | | | | (.761) | | | | |
| INT | .307 | (.812) | | | | .319 | (.882) | | | | .303 | (.766) | | | |
| PEOU | .184 | .135 | (.758) | | | .343 | .248 | (.753) | | | .046 | .096 | (.720) | | |
| ATT | .324 | .473 | .134 | (.766) | | .423 | .409 | .339 | (.719) | | .271 | .528 | .133 | (.785) | |

*Notes:* FL= factor loading; CR = composite reliability; AVE = average variance explained; MSV = maximum shared variance; ASV = average shared variance; PU = perceived usefulness; INT = intention to use; PEOU = perceived ease of use; PU = perceived usefulness
[a] Values on the diagonal of correlation matrices represents √AVE

Composite reliability and average variance extracted were used to estimate the reliability and convergent validity of the factors. Hair et al (2010)suggest that the CR value should be greater than 0.6 and that the AVE should be greater than 0.5. As can be shown in Table 6, the average variance extracted (AVE) within the British sample were all above 0.533 and above 0.758 for CR, whereas within the Lebanese sample, the AVE was above 0.518 and 0.759 for CR. Therefore, all factors have adequate reliability and convergent validity. Additionally, the total AVE of the average value of variables used for the research model for both samples is larger than their correlation value (Fornell and Larcker, 1981); therefore discriminant validity was also established for both samples.

### 3.3    Measurement Invariance

The procedure for measurement invariance mirrors the approach taken by Teo et al (2009), which involves producing a configurally invariant model during multi-group analysis in AMOS and adding increasingly strict invariance constraints. When good model fit is achieved, despite the increasing number of constraints, the model is deemed to be invariant across the two groups.

However, there is not much disciplinary consensus about which values for which fit indices indicate adequate fit. Traditionally, fit would be determined using the minimum fit function $\chi^2$. However, the $\chi^2$ may not be appropriate at large sample sizes because it can be overly sensitive to small differences (Hu and Bentler, 1999). The ratio of the $\chi^2$ static to its degree of freedom ($\chi^2/df$) is often used, where the value should be less than 3 to indicate a good fit of the data (Carmines and McIver, 1981). Many researchers have also suggested other fit indices to indicate acceptable fit (Hair *et al.*, 2010; Anderson and Gerbing, 1988; Steenkamp and Baumgartner, 1998). This study used the Non-Normed Fit Index (NNFI); Root Mean Square Residuals (RMSR); Comparative Fit Index (CFI); and the Root Mean Square Error of

Approximation (RMSEA) to evaluate the model fit of the both model. The Akaike Information Criterion (AIC) is also listed to provide readers with a relative indication of comparative model quality. As can be shown blow in Tables 7 and 8, and the following sections, the questionnaire items achieve partial measurement invariance in both the Lebanese and British contexts.

**Table 7**
Fit Indices for Invariance Tests

| Invariance Test | χ² | df | χ²/df | p | NNFI | CFI | SRMR | RMSEA | AIC |
|---|---|---|---|---|---|---|---|---|---|
| British Sample | 99.363 | 48 | 2.070 | .000 | .932 | .950 | .0550 | .073 (.053, .093) | 159.363 |
| Lebanese Sample | 58.456 | 48 | 1.218 | .143 | .985 | .989 | .0453 | .031 (.000, .055) | 252.243 |
| Baseline Model (Pooled) | 86.133 | 48 | 1.794 | .001 | .973 | .981 | .0419 | .043 (.028, .057) | 146.133 |
| Full Configural Invariance | 157.837 | 96 | 1.644 | .000 | .958 | .969 | .0453 | .038 (.027, .049) | 277.837 |
| Full Metric Invariance | 167.513 | 104 | 1.611 | .000 | .960 | .968 | .0448 | .037 (.027, .048) | 271.513 |
| Full Scalar Invariance | 246.915 | 112 | 2.205 | .000 | .921 | .933 | .0470 | .053 (.044, .062) | 382.915 |
| Partial Scalar Invariance | 174.578 | 108 | 1.616 | .000 | .960 | .967 | .0450 | .038 (.027, .048) | 318.915 |
| Full Factorial Invariance | 198.394 | 112 | 1.771 | .000 | .949 | .957 | .0505 | .042 (.032, .052) | 334.394 |
| Partial Factorial Invariance | 175.932 | 110 | 1.599 | .000 | .961 | .967 | .0454 | .037 (.027, .047) | 315.932 |
| Final Structural Model | 176.019 | 110 | 1.600 | .000 | .961 | .967 | .0451 | .037 (.027, .047) | 316.019 |

*Notes:* df = degrees of freedom, NNFI = non-normed fit index; CFI = comparative fit index; SRMR = standardised root mean square residual; RMSEA = root mean square error of approximation; AIC = akaike information criterion

**Table 8**
Results of χ² Difference Tests

| Model Comparison | Δ χ² | Δ df | p | Δ CFI | Decision |
|---|---|---|---|---|---|
| Test for Full Metric Invariance (M1 + M2) | 9.676 | 8 | .289 | -.001 | Supported |
| Test for Full Scalar Invariance (M2 + M3) | 79.402 | 8 | .000 | -.035 | Reject |
| Test for Partial Scalar Invariance (M2 + M4) | 7.065 | 4 | .132 | -.001 | Supported |
| Fest for Full Factorial Invariance (M4 + M5) | 23.816 | 4 | .000 | -.010 | Reject |
| Test for Partial Factorial Invariance (M4 + M6) | 1.353 | 2 | .508 | .000 | Supported |

*Notes:* Δ χ² = difference in chi-square values; Δ df = difference in degrees of freedom; Δ CFI = difference in comparative fit index value.

### 3.3.1 Configural Invariance
Configural invariance is satisfied when the basic model structure, such as the relationships between indicators and latent factors, is invariant across the groups. This initial baseline has no between-group invariance constraints, so differences may still exist in factor loadings, intercepts and variances, but it provides a basis for comparison as such constraints are added. It is, however, a critical step because if the data does not support identical patterns of fixed and non-fixed parameters, then the data will not support more restrictive models (Bollen, 1989).

### 3.3.2 Metric Invariance
Metric invariance supposes that the distance between item-responses (e.g. agree, strongly agree) in a scale represent an equal level of change in latent factor true score across independent samples. To test whether metric invariance is supported by the data, the model in AMOS was constrained, such that the factor loadings (also called the factor loading coefficients) were equal for both groups. Since the constrained model is nested within the model that tested for configural invariance, the results of a $\chi^2$ difference test were examined. A model that achieves metric invariance would have both, good fit to the data in addition to a non-significant difference to the previous model. However, while $\chi^2$ is widely used, researchers suggest that other fit indices, such as CFI, should also be used to evaluate model fit where a difference greater than 0.1 indicates a significant difference (Anderson and Gerbing, 1988; Steenkamp and Baumgartner, 1998; Hair *et al.*, 2010). As can be seen in Table 8 above, the non-significant $\chi^2$ difference (p = .289) and CFI difference (ΔCFI = -.010) indicates that full metric invariance has been achieved.

### 3.3.3 Scalar Invariance

Even though items may be metrically invariant, they may not be scalar invariant. This means that the intercept value (as in regression) may be different across the two groups. Such a result would suggest that a member of one group who responds with 'agree' may actually be indicating a different level of agreement compared to a member of another group who also responds with 'agree'. As can be seen in Table 8 above, the significant $\chi^2$ difference (p < .001) and CFI difference ($\Delta$CFI = -.350) indicates that full scalar invariance was not achieved. However, while some items were not invariant, at least one item on each factor was scalar invariant. Testing the model with fewer constraints, therefore, suggests that partial scalar invariance was established ($\chi^2$ (4) = 7.065, p = .132, $\Delta$CFI = -.001).

### 3.3.4 Factorial Invariance

Factorial invariance suggests that the two groups are homogenous in terms of factor structure; therefore, the variance for each factor should be identical across the two groups. As can be seen in the tables 7 and 8, full factorial invariance was not achieved ($\chi^2$ (4) = 23.816, p = .000, $\Delta$CFI = .010), but partial invariance was achieved ($\chi^2$ (2) = 1.353, p = .508, $\Delta$CFI = .000). This suggests that some factors were invariant across the two samples, suggesting there would be no significant differences, however there were likely to be differences in some of the factors.

## 3.4 Hypothesis Testing

Several hypotheses were stated in section 1.3, which are addressed here. Note, that $H_4$ was addressed in 2.1, where the significant Pearson's chi-squared statistic showed that the 36% of those in the British sample had experience with RSS, as opposed to just 21% in the Lebanese sample. The total sample size ($n$ = 437) was checked using Westland's (2010) calculator for adequacy, which revealed the proposed size was 136% of the minimum advised size for hypothesis testing on the model parameters specified. A series of z-tests were applied to the causal relationships in the structural model to identify differences between the British and Lebanese samples, as shown in Table 9.

**Table 9**
Hypothesis Testing Results for Applicability of the Technology Acceptance Model

|  |  | British Sample | | Lebanese Sample | | z-Tests | | |
|---|---|---|---|---|---|---|---|---|
|  |  | r | p | r | p | z | p | Result |
| H1 | PU <--- PEOU | 0.372 | .000 | 0.095 | .296 | 2.449 | .039 | Not Supported |
| H2a | ATT <--- PU | 0.342 | .000 | 0.338 | .000 | 0.049 | .796 | Supported |
| H2b | ATT <--- PEOU | 0.198 | .000 | 0.220 | .024 | -0.191 | .783 | Supported |
| H3a | INT <--- PU | 0.210 | .017 | 0.203 | .000 | 0.062 | .796 | Supported |
| H3b | INT <--- ATT | 0.577 | .000 | 0.521 | .000 | 0.476 | .712 | Supported |

The results support $H_{2a}$, $H_{2b}$, $H_{3a}$ and $H_{3b}$. All of the expected paths were significant and the results of z-tests comparing the correlation coefficients for differences were non-significant. However, $H_1$ was not supported. The relationship between PU and PEOU in the Lebanese sample was non-significant, and the result of the z-test comparing the path loading to the British sample was significant. This suggests that the relationship between these constructs differ, however the TAM model does predict that PU and PEOU are affected by external variables. Thus, hinting at some cultural influence.

**Table 10**
Mean Differences Between British and Lebanese Samples

|  |  | British Sample | | Lebanese Sample | | t-Tests | | | | |
|---|---|---|---|---|---|---|---|---|---|---|
|  |  | M | SD | M | SD | t | $z_{(\Delta M)}$ | p | d | Result |
| H5a | ATT | 3.104 | 0.539 | 2.910 | 0.671 | -3.390 | -0.311 | .001 | -0.32 | Supported |
| H5b | PU | 3.141 | 0.590 | 3.135 | 0.550 | -0.101 | -0.010 | .920 | n.s | Not Supported |
| H5c | PEOU | 3.066 | 0.622 | 3.410 | 0.486 | 6.556 | 0.608 | .000 | 0.62 | Not Supported |
| H5d | INT | 3.512 | 0.753 | 3.358 | 0.728 | -2.196 | -0.207 | .029 | -0.21 | Supported |

*Notes:* M = mean; SD = standard deviation; $z_{(\Delta x)}$ = standardised (z-)score difference; d = cohen's d statistic

The data support $H_{5a}$, and $H_{5d}$. The independent sample t-tests showed that the means of the factor scores for ATT and INT were significantly larger in the British sample, as expected by the support for $H_4$. However, these differences were "small" by Cohen's conventions on effect size. However, $H_{5b}$ was not supported. There was no significant difference between British and Lebanese students' perception that RSS can be useful, with both means being larger than 3 indicating agreement. Interestingly, however, $H_{5c}$ was not supported but a significant different was found. The Lebanese sample had much higher scores for PEOU on average, with "medium" effect. This suggests something about the culture that may help students to believe that that they can easily use, or learn to use, RSS feeds.

## 4 Discussion and Conclusion

The overall aim of this study was to explore the factors affecting students' intention to adopt and use RSS feeds in higher education setting and investigate whether there are differences among the predictors of technology acceptance between British and Lebanese students. Our result supports the ability of TAM to be a useful theoretical framework for better understanding the student's behavioural intention to use RSS feeds in education in both Lebanon and Britain. In general, the results show that the majority of participants in the British and Lebanese sample express positive responses to the constructs being measured by the questionnaire. This means that both British and Lebanese students are willing to embrace RSS feeds as part of their repertoire of learning opportunities. More specifically, the results of the proposed model show that all the predictors (PEOU, PU, ATT) were found to be significant determinants of behavioural intentions to use RSS feeds for both samples.

Initially, invariance tests were conducted to remove variant items from the research instrument. Then, Fischer's $X^2$ statistic was applied as a 'sanity test' before proceeding to further multi-group analysis, which assessed configural, metric, scalar, and factorial invariance. Such testing was conducted to ensure that the psychometric properties of the measure used to explore the two samples had equivalent structure and properties. The results suggest that the TAM questionnaire items included in the analysis were robust across the two cultures studied, due to the evidence of adequate configural, metric and (partial) scalar invariance found. This suggests that the factor structure and related factor loadings for each item in the measure were equivalent. Therefore, it was appropriate for use in comparing the two cultures. As anticipated, however, full factorial invariance was not established. Thus, the two samples were not homogenous, suggesting at least some significant differences in the relationships between constructs.

Overall, the proposed measurement model achieved acceptable fit, explaining total cumulative variances of 73% for the British sample and 71% for the Lebanese sample. However, further exploration indicated that the differences between the Lebanese and British students are greater than the similarities. More specifically, significant differences were detected in terms of ATT, PEOU and INT, whereas no differences were detected in terms of PU. Furthermore, the relationship between PU and PEOU was significantly different. Nevertheless, the proposed structural model had similar predictive qualities, with the squared multiple correlation of INT being 39% for the Lebanese sample and 24% for the British sample. Despite these differences, the results indicate that both Lebanese and British students would most likely use RSS feeds in their learning process, but reports from the data collection suggested many students were not aware of the feature was available in the virtual learning environment until being presented the questionnaire. Therefore, it is recommend that educators spread awareness of the feature, emphasise the usefulness of the feature and how it can be used to benefit learning in order to encourage intention for use.

Generally speaking, we do not know if technology that has been developed in one location would be perceived in similar ways in different locations. By establishing a cross-cultural validation of a model, this would not only achieve greater validity of research instruments, but also allow meaningful comparisons and analyses between and within samples to be made. This research moves in this direction.
As with any research, this study has some limitations. Firstly, we did not integrate the cultural variables within the model and assumed Hofstede's findings to be true. Future research might investigate the direct or moderating effect of culture on students' perceptions towards using RSS. Further research is also needed to specifically examine the influences of other individual characteristics such as gender, age, educational level and experience on the acceptance and use of RSS in education. Secondly, data were collected from students using a convenience sampling technique and thus should not necessarily be considered representative of the population. Therefore, generalization of these findings should be treated with caution. Nevertheless, as a practice, it is acceptable as a first step for further exploration because it is the position of the authors that, from a measurement perspective, a scale found to be variant and have problems using a non-probability sample from a small local sampling frame is unlikely to be invariant using a probability sample in a national sampling frame. Future research should examine the students' intention to use RSS feeds in environments where the use of e-learning tools is voluntary as it was found that this variable has a big influence on students' perception towards using technology (Venkatesh *et al.*, 2003), with different user groups (e.g. students with physical impairments) and different organizational contexts (e.g. high schools or public institutions) to explore the validity of the model in different contexts.